\documentclass[10pt]{article}

\usepackage{amsmath}
\usepackage{amssymb}
\usepackage{graphicx}
\usepackage{cite}
\usepackage{color} 

\usepackage[labelfont=bf,labelsep=period,justification=raggedright]{caption}
\makeatletter
\renewcommand{\@biblabel}[1]{\quad#1.}
\makeatother
\date{}

\begin{document}

\begin{flushleft}
{\Large
\textbf{Simulation of a Microfluidic Gradient Generator using Lattice Boltzmann Methods}
}
\\
Simon Tanaka$^{1,2}$, 
Dagmar Iber$^{1,2}$
\\
\bf{1} Department for Biosystems Science and Engineering, ETH Zurich, Mattenstrasse 26, Basel,
Switzerland, +41 61 387 32 10 (phone), +41 61 387 31 94 (fax)
\\
\bf{2} Swiss Institute of Bioinformatics, Basel, Switzerland
\\
$\ast$ E-mail: Corresponding author dagmar.iber@bsse.ethz.ch
\end{flushleft}

\section*{Abstract}

Microfluidics provides a powerful and versatile technology to accurately control spatial and temporal conditions for cell culturing and can therefore be used to study cellular responses to gradients.
Here we use Lattice Boltzmann methods (LBM) to solve both the Navier-Stokes equation (NSE) for the fluid and the coupled convection-diffusion equation (CDE) for the compounds that form the diffusion-based gradient.
The design of a microfluidic chamber for diffusion-based gradients must avoid flow through the cell chamber.
This can be achieved by alternately opening the source and the sink channels.
The fast toggling of microfluidic valves requires switching between different boundary conditions.
We demonstrate that the LBM is a powerful method for handling complex geometries, high P\'eclet number conditions, discontinuities in the boundary conditions, and multiphysics coupling.

\section*{Introduction}

The spatial organization of complex organisms requires cells to read out concentration differences and respond accordingly. Graded responses have been documented in different contexts, including developmental processes and immunological responses. To study these processes in greater detail it is important to study cells in a well-controlled setting where concentration gradients can be applied. Microfluidics provides a powerful and versatile technology to accurately control spatial and temporal conditions. A standard layout of a microfluidic set-up is shown in Figure \ref{fig:setup}. The chamber is connected to two channels that act as source and sink of the compound of interest. The compound is transported through these channels mainly by advection. Within the chamber, however, flow must be avoided as the resulting hydrodynamic stress would impact on the cells. Transport of the compound in the chamber must therefore be diffusion-dominated. The transition from advection- to diffusion-dominated regime at the right place is a challenging engineering problem. In the design phase, the use of adequate simulation techniques can provide insight into the governing mechanisms.\\

To describe the process, we need to simulate both the fluid dynamcis and the distribution of the compound in the fluid. An incompressible Newtonian fluid is described by the Navier-Stokes equation (NSE):
\begin{equation}
 \rho \left( \frac{\partial \boldsymbol{u}}{\partial t} + \left( \boldsymbol{u}\cdot \nabla\right) \boldsymbol{u} \right) = 
  -\nabla p + \mu \nabla^{2} \boldsymbol{u} + \boldsymbol{f}
\end{equation}
where $\boldsymbol{u}$ denotes the velocity field, $\rho$ the fluid density, $\mu$ the dynamic viscosity, and $\boldsymbol{f}$ an external body force, which is zero in our case.
The passive advection and  diffusion of a diluted compound is described by the advection-diffusion equation:
\begin{equation}\label{eq:cde}
 \frac{\partial C}{\partial t} + \nabla \cdot \left(C \boldsymbol{u} \right) = D\nabla^{2}C + R
\end{equation}
with the diffusion coefficient $D$, and the local reaction term $R$.\\

Several numerical schemes have been used to simulate fluid flow, including the Lattice Boltzmann method (LBM) \cite{Chen1998}. It has been shown that the Navier-Stokes equation is recovered in the nearly incompressible hydrodynamic limit. The LBM was also successfully applied to solve diffusion \cite{Wolf-Gladrow1995} and advection-diffusion equations \cite{PonceDawson1993}. Furthermore, the LBM has been shown to be capable to simulate reaction-diffusion problems by applying it to Turing type mechanisms \cite{Blaak2000,Li2001,Ayodele2011}.
The coupling of a passively advected and diffusing scalar to a fluid was shown in \cite{Inamuro2002, Guo2002b}.

The Lattice Boltzmann method has previously been used to simulate a non-flow-free microfluidic gradient generator \cite{Hu2011}, whose design was inspired by \cite{Dertinger2001}. However, since the microfluidic gradient generator is not flow-free, it would not be suitable for experiments with cells that are sensitive to flow. Also, the P\'eclet number in these simulations is about fivefold lower than in the experimental data that is used to validate the simulations, and well within the numerically non-critical regime of the standard Lattice Boltzmann methods.   

Recently, a microfluidic setup was published that prevents flow through the cell chamber by using alternate flushing of the channels \cite{Frank2013}. We use LBM-based simulations to computationally explore a similar design. Because of its algorithmic locality, complex geometries and boundary conditions (such as the fast switching of boundary conditions to imitate the valve dynamics) can easily be integrated into the LBM. In the following we will describe the LBM-based simulations of gradient formation in a microfluidic chamber.

\section*{Methods}

\subsection*{Experimental Layout and Conditions}

\begin{figure}[!ht]
\begin{center}
\includegraphics[width=0.8\textwidth]{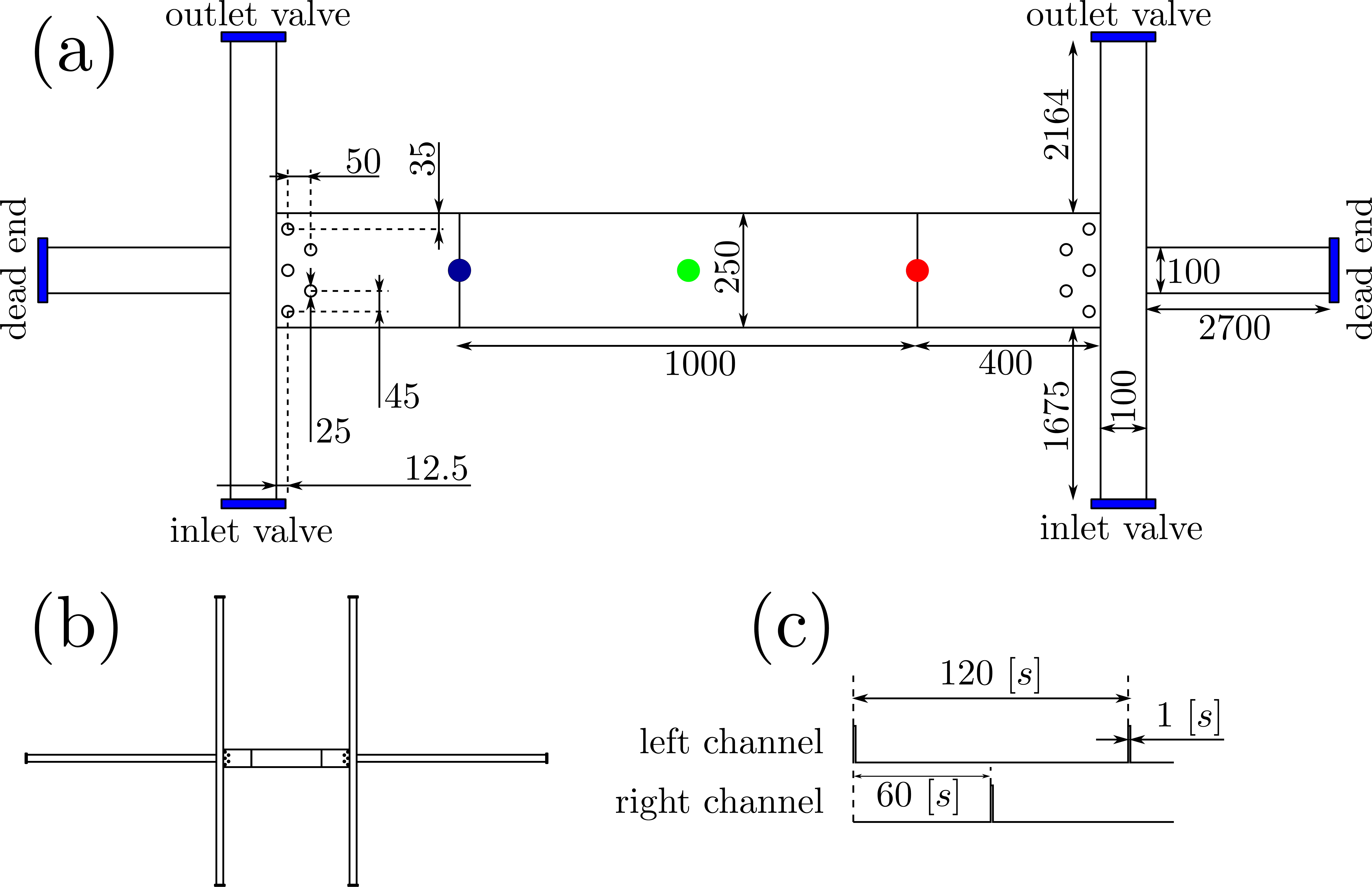}
\end{center}
\caption{}
\label{fig:setup}
\end{figure}

The implemented layout of the microfluidic chip is shown in Figure \ref{fig:setup} (a). It consists of two parallel channels of width $100\,\left[\mu m\right]$ and length $1000\,\left[\mu m\right]$, whose inlet and outlet are controlled by microfluidic valves. The two channels are connected by the diffusion chamber, consisting of a pre-chamber region, and the actual cell chamber of size $1000\,\left[\mu m\right]$ by $250\,\left[\mu m\right]$. A fence consisting of five cylindrical elements disturbs the flow and thus reduces its advective influence in the chamber. Additional dead-end channels, which are used for cell handling, are explicitly modeled since they act as reservoirs. The height of the chamber is between $40\,\left[\mu m\right]$ and $100\,\left[\mu m\right]$, such that the processes are well approximated by a two dimensional simulation.

The distance from the valves to the chambers may be larger in reality (cf. Figure \ref{fig:setup} (b)); however, they can be pruned such that they are in the range of the effective diffusion length $2\sqrt{DT}$, defined by the diffusion coefficient $D$ and the time period $T$ between two consecutive flushings.

Every two minutes, the left channel opens both its inlet and outlet valve for one second.
As shown in Figure \ref{fig:setup} (c), the sequence for the right channel is shifted by one minute, such that the chip is completely closed for 59 seconds.
The technique to avoid flow through the cell chamber using alternate flushing of the channels has been described in \cite{Frank2013}.

In the beginning, the left channel is flushed with medium with a compound of interest, and the right channel with plain medium. Therefore a concentration gradient emerges across the chamber. After two hours, the fluids are interchanged for another two hours, enforcing an inversion of the gradient. 

Upon opening the valve, the flow is virtually immediately fully developed due to the very low Reynolds number and the incompressibility. The high maximal flow speed in the channel $U^{\text{phys}}=5000\,\left[\mu m/s\right]$ and the low diffusion coefficient of the compound $D^{\text{phys}}=122\,\left[\mu m^{2}/s\right]$ result in an advection-dominated transport in the channels. Using a characteristic length scale $L=100\,\left[\mu m\right]$ (the channel width), the P\'eclet number $Pe= UL/D\approx 4100$ can be computed. While entering the diffusion chamber (of after closing all valves), the advective transport reduces to zero and the diffusive transport dominates.

A summary of the used physical parameter values can be found in Table \ref{tab:fluidparameters} and \ref{tab:advectiondiffusionparameters}.

\subsection*{Numerical Methods}

For the fluid, we employed the standard D2Q9 solver proposed in \cite{Chen1998}.
The D2Q9 lattice is defined by the lattice directions $\boldsymbol{e}_{i}$:
\begin{equation}\label{eq:lattice}
 \boldsymbol{e}_{i} = \begin{cases}
		      \left[0,0\right], & \mbox{for } i=0 \\
		      \left[\cos\left(\pi\left(i-1\right)/2\right), \sin\left(\pi\left(i-1\right)/2\right)  \right], & \mbox{for } i=\{1,2,3,4\} \\
		      \sqrt{2} \left[\cos\left(\pi\left(i-9/2\right)/2\right), \sin\left(\pi\left(i-9/2\right)/2\right)  \right], & \mbox{for } i=\{5,6,7,8\} \\
		      \end{cases}
\end{equation}

The lattice Bhatnagar-Gross-Krook (BGK) equation, the equilibrium distribution function $f_{i}^{eq}$ in the $i$th direction, and the corresponding weights $w_{i}$ are taken as:
\begin{equation}\label{eq:lbequation}
 f_{i} \left( \boldsymbol{x} + c \boldsymbol{e_{i}} \Delta t, t+\Delta t\right) - f_{i} \left( \boldsymbol{x},t \right) =
			-\frac{1}{\tau_{f}} \left[ f_{i}\left(\boldsymbol{x},t\right) - f_{i}^{eq}\left( \boldsymbol{x},t \right) \right]
\end{equation}

\begin{equation}
 f_{i}^{eq} = \rho w_{i} \left[ 1+3 \boldsymbol{e}_{i}\cdot\boldsymbol{u} + \frac{9}{2}\left(\boldsymbol{e}_{i}\cdot\boldsymbol{u}\right)^{2} - \frac{3}{2} u^{2} \right]
\end{equation}

\begin{equation}
 w_{i} = \begin{cases}
		      4/9, & \mbox{for } i=0 \\
		      1/9, & \mbox{for } i=\{1,2,3,4\} \\
		      1/36, & \mbox{for } i=\{5,6,7,8\} \\
		      \end{cases}
\end{equation}
where $\Delta t$ denotes the time step, $\Delta x$ the spatial discretization and $c=\Delta x/\Delta t$ the lattice speed.
The simplest choice to guarantee consistency with the lattice (Eq. (\ref{eq:lattice})) is $\Delta t=\Delta x=c=1$.
The relaxation time $\tau_{f}$ is related to the kinematic viscositiy $\nu = c_{s}^{2}\left(\tau_{f} - 1/2\right)$.
For isothermal flow, the speed of sound is defined as $c_{s} = 1/\sqrt{3}$.

Equation (\ref{eq:lbequation}) implies a two step algorithm: first, local collision relaxes the populations towards the local
equilibrium (right hand side), and second, the populations perform a free flight to the next lattice point (left hand side).
The density, momentum density and pressure are computed as:
\begin{equation}\label{eq:moments}
 \rho = \sum_{i=0}^{9} f_{i} \hspace{1cm} \rho \boldsymbol{u} = \sum_{i=0}^{9} f_{i} \boldsymbol{e}_{i} \hspace{1cm} p=\rho c_{s}^{2}
\end{equation}

For solving the advection-diffusion equation of a compound $C$ (Eq. (\ref{eq:cde})), a multi-distribution function (MDF) approach was used.
It was first used in \cite{BARTOLONI1993} to model the temperature field of a thermal flow (Boussinesq approximation) as a passively advected scalar field,
and further improved in \cite{Guo2002b}.
In contrary to the fluid solver, \cite{Guo2002b} use a D2Q4 lattice and stencil.
The lattice BGK equation, the equilibrium distribution and the zeroth order moment read:
\begin{equation}
 g_{i} \left( \boldsymbol{x} + c \boldsymbol{e_{i}} \Delta t, t+\Delta t\right) - g_{i} \left( \boldsymbol{x},t \right) =
			-\frac{1}{\tau_{g}} \left[ g_{i}\left(\boldsymbol{x},t\right) - g_{i}^{eq}\left( \boldsymbol{x},t \right) \right]
\end{equation}

\begin{equation}\label{eq:cdeequilibrium}
 g_{i}^{eq} = \frac{C}{4}\left[1 + 2\frac{\boldsymbol{e}_{i} \cdot \boldsymbol{u}}{c}\right]	\hspace{1cm}\mbox{with}\hspace{1cm}	C = \sum_{i=1}^{4} g_{i}
\end{equation}
where $g_{i}$ denote the particle distribution functions.
The velocity field $\boldsymbol{u}$ is transferred from the fluid solver (Eq. (\ref{eq:moments})).
The relaxation time $\tau_{g}$ is related to the diffusion coefficient as $D = (2 \tau_{g}-1)/4$.

Note that all variables are measured in the LB units $\delta_t$ and $\delta_x$.
The conversion to physical quantities is described below.
\subsection*{Boundary Conditions}

For the wall boundary condition of the fluid, the missing incoming populations are approximated by
equilibrium distributions.
The momentum needed to compute the equilibrium is spatially first order interpolated between the fluid in direction of the missing population, and the known zero momentum at the wall.
The density is spatially first order extrapolated from the fluid.

The pressure boundary condition for the fluid is realized by applying do-nothing, corresponding to zeroth order
extrapolation in time.
Then all popluations are rescaled such that the desired pressure is obtained.
The velocity field is not affected, since the rescaling factor $\gamma$ cancels:
\begin{equation}
\boldsymbol{u}\left(\boldsymbol{x},t\right) = \frac{\sum_{i} \gamma \boldsymbol{e}_{i} f_{i}\left(\boldsymbol{x},t\right)}{\sum_{i} \gamma f_{i}\left(\boldsymbol{x},t\right)}
\end{equation}
This technique was used in \cite{Zhang2006a}, although for specifying the pressure difference in periodically closed channels.

The same approach is used for the Dirichlet boundary condition of the diffusing compound at the channel inlet.
Since the prescribed-density boundary condition cannot be applied to the outlet, the well-known do-nothing boundary condition is applied \cite{HEYWOOD1996}.
For the no-flux boundary condition, the boundary condition presented in \cite{Guo2002a,Guo2002b} is used.

\subsection*{Choice of Parameter Values and Conversion}

For both the fluid and the advection-diffusion simulations, the flow channels are resolved by $20$ lattice points.
The lattice spacing thus derives as $\delta_{x}=100\,\left[\mu m\right]/20 = 5\,\left[\mu m\right]$.

For the fluid simulation, the dimensionless Reynolds-number can be used to determine the missing LB parameters: 
\begin{equation}
 Re = \frac{L^{\text{phys}} U^{\text{phys}}}{\nu^{\text{phys}}} =  \frac{L^{\text{LB}} U^{\text{LB}}}{\nu^{\text{LB}}} = 0.5
\end{equation}

With the choice of a numerically reasonable LB kinematic viscosity $\nu^{\text{LB}}=0.05$, the maximal fluid speed $U^{\text{LB}}$ can be computed (cf. Table \ref{tab:fluidparameters}).
The LB time unit can be computed by equating the physical and LB maximal fluid speed, which results in $\delta_{t}=1.25\times 10^{-6}\,\left[s\right]$
The pressure difference is chosen such that $U^{\text{LB}}$ is achieved with less than $1\%$ error.

The procedure for the advection-diffusion simulation is similar.
In order to allow for large time steps, the  maximal fluid speed is chosen as $U^{\text{LB}}=0.5$, which is close to the speed of sound $c_{s}=1/\sqrt{3}$. REF
Again, the LB time unit $\delta_{t}=5\times 10^{-4}\,\left[s\right]$ is determined by comparing the maximal fluid speeds.
To use the fluid field from the fluid simulation in the advection-diffusion simulation, the field was scaled according to the change in the LB time unit.
By equating the dimensionless P\'eclet number
\begin{equation}
 Pe = \frac{L^{\text{phys}} U^{\text{phys}}}{D^{\text{phys}}} =  \frac{L^{\text{LB}} U^{\text{LB}}}{D^{\text{LB}}} \approx 4.1 \times 10^{3} \, \left[1\right]
\end{equation}
the diffusion coefficient $D^{\text{LB}}$ can be computed (cf. Table \ref{tab:advectiondiffusionparameters}).

\section*{Results}

The diffusing compound is passively advected by the fluid, but does not feed back on the fluid dynamics. The fluid dynamics can therefore be solved independently. After having computed the velocity fields, they can be used as an input to solve the advection-diffusion equation of the compound. The results of these steps are analyzed and discussed in the following.
\newline

\begin{figure}[!ht]
\begin{center}
\includegraphics[width=0.8\textwidth]{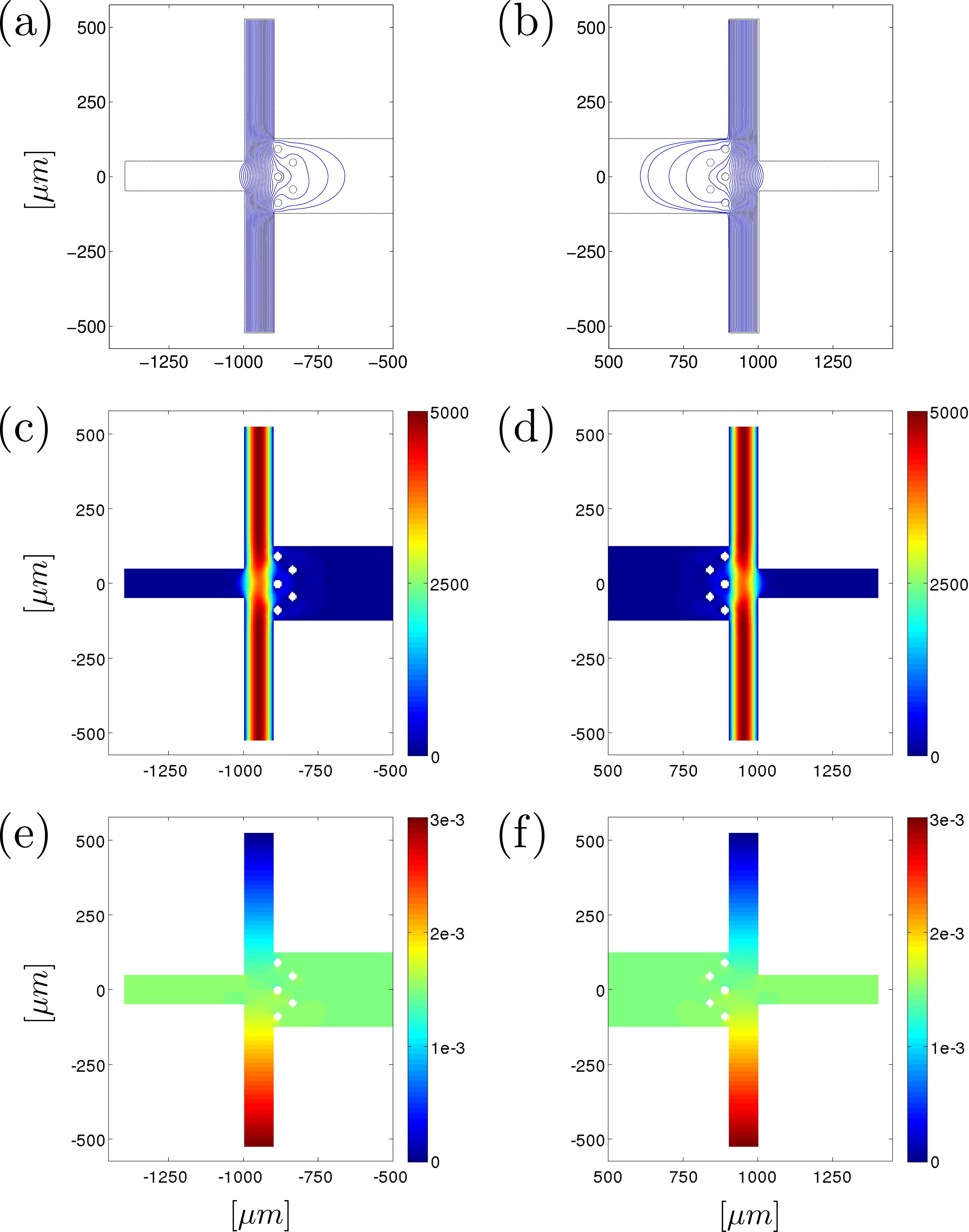}
\end{center}
\caption{}
\label{fig:flowfield}
\end{figure}

In a first step, only the fluid dynamics is solved and analyzed. 
The geometry as shown in Figure \ref{fig:setup} (a) was implemented using the well known D2Q9 BGK LB scheme.
Upon opening the valves, a pressure boundary condition with prescribed pressure was applied.
The fluid in the domain, initially at rest, is brought to a steady state flow field in a short time span in a weakly compressible manner.
The streamlines for opened left and right channel are shown in Figure \ref{fig:flowfield} (a) and (b), respectively.
Although the majority of streamlines follows a straight line, particles close to the boundaries flow around the fence and into the chamber.
It can be observed that the direction of flow is locally reversed.
However, the magnitude of the velocity is negligible in the chamber itself (Figure \ref{fig:flowfield} (c-d)), which is an important requirement for culturing cells.
In the channel, the flow approximates a Poiseuille flow velocity profile, and corruptions as a consequence of the inlet and outlet boundary conditions are negligible.
The fluid pressure, shown in Figure \ref{fig:flowfield} (e-f), drops, according to the Poiseuille solution, linearly in the channels.
Close to the chamber, the effective flow cross-section is increased, leading to a lower pressure gradient.
The pressure field in the cell chamber is homogeneous.
The chosen parameter values for the fluid dynamics simulation are given in Table \ref{tab:fluidparameters}.
\newline 

Since the flow virtually immediately reaches steady state (after about $0.025\,\left[s\right]$), and because the time step for the fluid solver has to be chosen considerably smaller (about tenfold) than the desired time step for the advection-diffusion solver, the velocity fields are pre-computed (as described in the aforegoing section), stored and then loaded into the advection-diffusion solver.
The velocity fields of three different conditions are needed: left channel open and right channel closed, left channel closed and right channel open, and all channels closed.
This approach prevents the advection-diffusion solver from solving the same fluid flow repeatedly.
\newline

In a second step, the advection-diffusion solver, which is largely consistent with the fluid solver, is implemented to solve Equation (\ref{eq:cde}).
Other than in the fluid solver, the velocity to compute the equilibrium distributions (Equation (\ref{eq:cdeequilibrium})) is taken from the pre-computed fluid velocity field.
The zeroth order moment (density) can be interpreted as the compound concentration.
In order to allow for maximal temporal step size, the fluid velocity has to be chosen as high as possible.
We chose $U^{\text{LB}}=0.5$ in order to comply with the limit defined by the speed of sound, $c_{s}$.
It has been shown that, for advection-diffusion applications, the advection velocity can be chosen comparably high \cite{Suga2006,Chopard2009}, although the error grows.
However, the choice of a new $U^{\text{LB}}=0.5$ (and thus a new time step $\delta_{t}$) requires the rescaling of the pre-computed velocity fields.

In order to simulate the opening and closing of the valves, the inlet and outlet boundary conditions are changed periodically for the advection-diffusion solver.
For closed valves, an ordinary wall (no-flux) boundary condition is applied.
An open inlet is realised by a prescribed-density boundary condition, which corresponds to the pressure boundary condition in the fluid solver.
The inlet density is set to $1$ or $0$, depending on whether the channel is fed with the concentrated or the plain medium.
\newline

\begin{figure}[!ht]
\begin{center}
\includegraphics[width=0.8\textwidth]{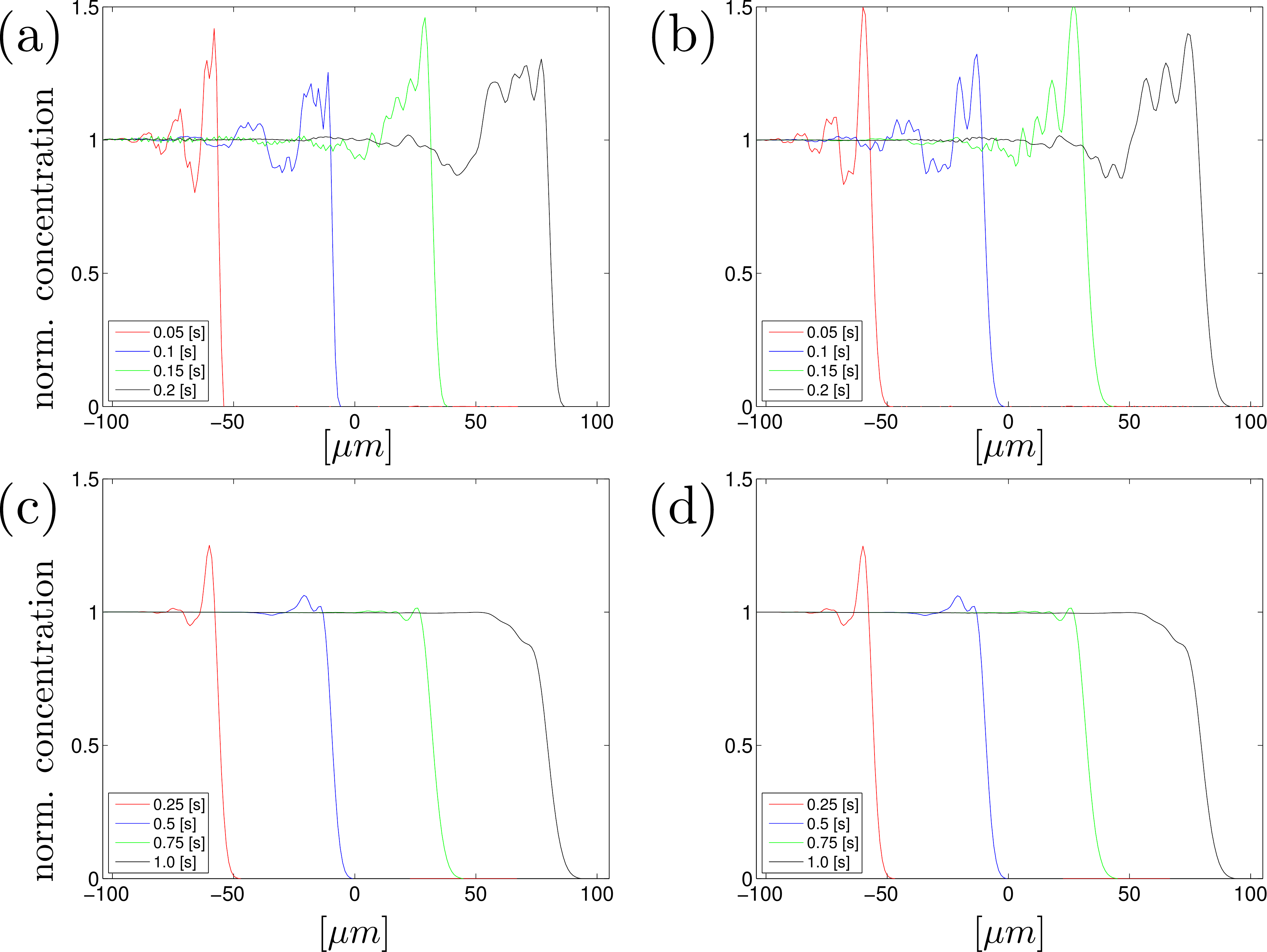}
\end{center}
\caption{}
\label{fig:shockfigure}
\end{figure}

When opening the channel carrying the concentrated medium for the first time, the advected concentration profile  has a step-like shape. The advection of discontinuities (or extremely sharp gradients) is numerically challenging.
To test the capability of the advection-diffusion solver to advect a step profile with very low diffusion, the initial opening of the left channel is considered.
At time $t=0$, the channel is at rest, and the compound concentration is zero.
Upon opening the channel, the fluid velocity immediately accelerates to the stationary (pre-computed) velocity field.
The compound concentration, being $1$ at the inlet boundary, is carried with the flow into the channel.
This situation is similar to the advection of a step profile, with two differences: firstly, the velocity is not constant perpendicular to the direction of flow; and, secondly, we have diffusive transport.
However, the P\'eclet number $Pe\approx 4100$ is very high, meaning that the advective transport dominates the diffusive transport, and the smoothing of the initially step-like concentration profile is minor.
Figure \ref{fig:shockfigure} (a) shows the concentration profile along the channel.
Severe numerical instabilities are induced close to the high gradient region, which do not decay with the front being advected.
A tenfold decrease of the time step $\delta_{t}$ does not affect the instabilites (cf. Figure ref{fig:shockfigure} (a)).
Decreasing, instead, the P\'eclet number to $Pe\approx 800$ by decreasing the flow speed to $U^{\text{phys}}=1000\,\left[\mu m/s\right]$ (and thus $U^{\text{LB}}=0.1\,\left[\delta_{x}/ \delta_{t}\right]$) leads to significant smoothing of the initially present numerical instabilities, as shown in Figure \ref{fig:shockfigure} (c).
Again, decreasing the time step tenfold does not lead to an improvement of the solution (cf. Figure \ref{fig:shockfigure} (d)).
However, since the numerical instabilities only form at the very first opening on a massive scale, and only on a minor scale at later openings, they are acceptable when considering the long time solution.
The chosen parameters for the step advection  test are summarized in Table \ref{tab:shockparameters}.
\newline

\begin{figure}[!ht]
\begin{center}
\includegraphics[width=0.8\textwidth]{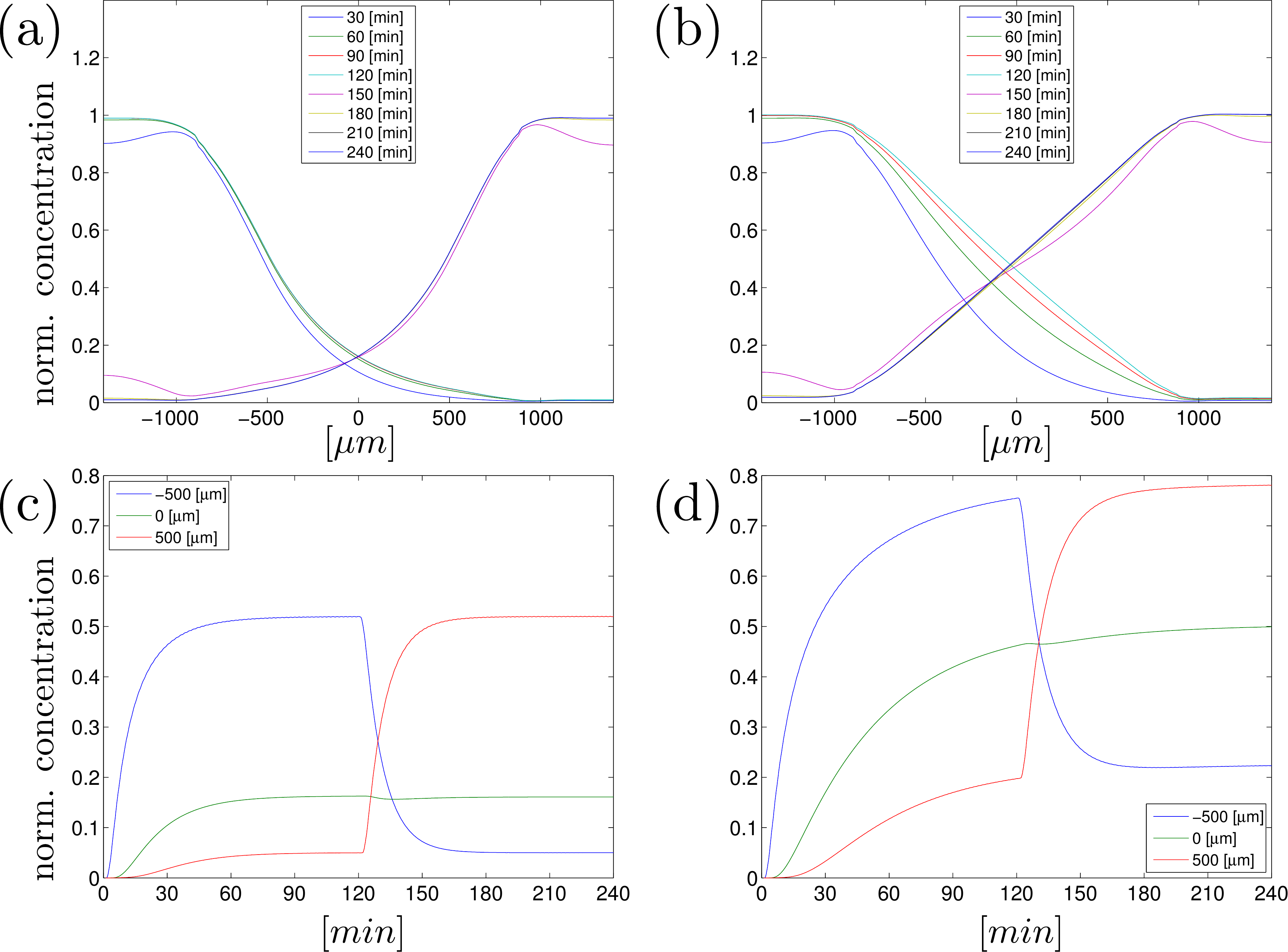}
\end{center}
\caption{}
\label{fig:concentrationfigure}
\end{figure}

As last step, the full microfluidic setup is simulated.
The left channel is fed with concentrated medium for two hours, leading to a gradient from the left to the right.
For another two hours, the gradient is reversed by changing the medium of the channels.
In order to test the effect of decay of the compound, the simulations are once carried out with, and once without degradation.
In the latter case, the gradient reaches almost its steady-state after $60\,\left[min\right]$ (shown in Figure \ref{fig:concentrationfigure} (a)). According to the theory, a non-linear steady-state concentration profile is obtained.
Thirty minutes after reversing the media, the gradient already reaches its mirrored steady-state profile.
However, the dead-end channels act as reservoirs, revealing the importance of their inclusion into the simulation.
When setting the compound degradation to zero, it takes approximately two hours to reach the linear steady-state concentration profile (shown in Figure \ref{fig:concentrationfigure} (b)). The reversion is much faster and the steady-state is already reached after one hour.
This can be explained by the fact that a particle only has to diffuse through half the chamber, whereas it had to diffuse through the entire chamber to form the initial gradient.
To analyze the dynamics of gradient formation, the time evolution of the concentration at fixed points (shown in Figure \ref{fig:setup} as blue, green and red points) is shown in Figure \ref{fig:concentrationfigure} (c)) for a decaying compound, and in \ref{fig:concentrationfigure} (d)) for a non-decaying compound.
Besides the time to reach steady-state, also the reached maximal concentrations in the cell chamber (from $-500\,\left[\mu m\right]$ to $500\,\left[\mu m\right]$) are considerably lower in the case when the compound decays.

\section*{Discussion}

In order to gain insight into the governing processes in a microfluidic gradient generator, and to support its design and development, computer simulations based on Lattice Boltzmann methods were developed.
We have shown that the LBM is capable to simulate both the fluid flow and the coupled advective and diffusive processes.
Although the high P\'eclet number regime ($Pe>4000$) leads to severe numerical instabilities when advecting steep gradients, we report that, on the long time scale, the LBM leads to stable solutions.
Using simulation it was shown that the presented microfluidic chip design is capable of forming and maintaining spatially and temporally controlled, diffusion based gradients.
The valve-switching strategy and the flow-disturbing fence reliably prevent the flow from entering the cell culture chamber.

The benefit of the availability of such a gradient generator is manifold.
Firstly, the concentration levels at points of interest can be accurately predicted and controlled.
With the steadily increasing demand for quantitative data, microfluidic gradient generators offer the possibility to efficiently conduct multiple experiments on one chip.
Secondly, the compound sources and sinks are spatially close to the experimental chamber and frequently renewed.
This allows to use biochemical compounds with very high (self-) degradation, since the diffusive transport only has to cover comparably small distances.

\section*{Acknowledgments}

The authors acknowledge funding from the SNF Sinergia grant ÓDevelopmental engineering of endochondral ossification from mesenchymal stem cellsÓ and from the SNF SystemsX RTD NeurostemX.

\bibliography{MyCollection}

\clearpage

\newpage
\section*{Figure Legends}
\paragraph{Figure \ref{fig:setup}: Simulation Setup.}
(a) The channel and chamber geometry (all numbers given in $\left[\mu m\right]$).
The cell chamber of interest, inbetween the two channels, has dimensions $1000\,\left[\mu m\right]$ by $250\,\left[\mu m\right]$.
The dead end valves are always closed.
(b) In reality the distance from the valves to the chamber would be longer,
but the channels are pruned in the simulation setup in order to save simulation time.
This has no impact on the solution since the diffusion length is smaller than the length of the pruned channels.
(c) The left and right channel are opened alternately for $1\,\left[s\right]$, respectively.
Between the flushings, all valves are closed for $119\,\left[s\right]$.

\paragraph{Figure \ref{fig:flowfield}: The flow field.}
(a-b) Streamlines when the left (a) or right (b) channel are open, respectively.
(c-d) The magnitude of velocity. In the channel, a Poiseuille-flow-like velocity profile is developed.
The units of the color code is $\left[\mu m/s\right]$.
(e-f) The pressure field drops linearly in the channels, and flattens in the chamber-transition-zone, where the effective flow-cross-section is increased. The color code is given in LB units.

\paragraph{Figure \ref{fig:shockfigure}: Propagation of a step profile.}
The concentration profiles along the midline of the left channel is shown, shortly after opening the valve the first time.
The top row shows simulations for a P\'eclet number $Pe\approx 4100$, and the bottom row for $Pe\approx 800$ after decreasing the fluid speed to $U^{\text{phys}}=1000\,\left[\mu m/s\right]$. In the second column, the temporal resolution is tenfold increased as compared to the first columns.
The chosen parameters are summarized in Table \ref{tab:shockparameters}.
(a) For low temporal resolution and high $Pe$ number, high numerical instabilities are triggered, which do not decrease as time passes.
(b) Increased temporal resolution does not mitigate the instabilities.
(c) At a five-fold lower $Pe$ number, numerical instabilities can be initially found close to the high gradient, but they diminish as the step profile propagates and smoothens.
(d) Higher temporal resolution does not lead to significantly different results.

\paragraph{Figure \ref{fig:concentrationfigure}: Evolution of Concentration.}
(a) The concentration profiles at different time points along the midline of the chamber.
The compound is decaying, leading to a non-linear steady-state profile.
(b) For a stable compound, a linear concentration profile is obtained in the chamber.
(c) Time evolution of the concentration at three different locations in the chamber (cf. Figure \ref{fig:setup}).
The steady state is reached quickly for a decaying compound, as opposed to a stable compound (d).
  
\clearpage
\section*{Figures}

\clearpage
\section*{Tables}

\begin{table}[!ht]
    \centering
    \caption{\bf{Parameter values for the fluid solver, both in physical and LB units.}}
    \begin{tabular}{|l|l|l|}
	\hline 
	 & physical units & LB units \\ 
	\hline
	\hline
	channel width & $L^{\text{phys}} = 100 \, \left[\mu m\right]$ & $L^{\text{LB}}=20 \, \left[\delta_{x}\right]$ \\ 
	\hline 
	simulation time & $T^{\text{phys}}=  0.125\, \left[s\right]$ & $T^{\text{LB}} = 10^{5} \, \left[\delta_{t}\right]$ \\ 
	\hline 
	maximal velocity & $U^{\text{phys}} = 5\times 10^{3} \, \left[\mu m/s\right]$ & $U^{\text{LB}} = 0.00125 \, [\delta_{x}/\delta_{t}]$ \\
	\hline
	kinematic viscosity & $\nu^{\text{phys}} = 10^{6} \, \left[\mu m^{2}/s\right]$ & $\nu^{\text{LB}}= 0.05 \left[\delta_{x}^{2}/\delta_{t}\right]$ \\
	\hline 
	\end{tabular} 
    \label{tab:fluidparameters}
\end{table}

\begin{table}[!ht]
    \centering
    \caption{\bf{Parameter values for the CDE solver, both in physical and LB units.}}
    \begin{tabular}{|l|l|l|}
	\hline 
	 & physical units & LB units \\ 
	\hline
	\hline
	channel width & $L^{\text{phys}} = 100 \, \left[\mu m\right]$ & $L^{\text{LB}}=20 \, \left[\delta_{x}\right]$ \\ 
	\hline 
	simulation time & $T^{\text{phys}}= 4 \, \left[h\right] = 14400 \, \left[s\right]$ & $T^{\text{LB}} = 28800000 \, \left[\delta_{t}\right]$ \\ 
	\hline 
	maximal velocity & $U^{\text{phys}} = 5\times 10^{3} \, \left[\mu m/s\right]$ & $U^{\text{LB}} = 0.5 \, [\delta_{x}/\delta_{t}]$ \\
	\hline 
	diffusion coefficient & $D^{\text{phys}} = 122 \, \left[\mu m^{2}/s\right]$ & $D^{\text{LB}} = 0.00244 \, \left[\delta_{x}^{2}/\delta_{t}\right]$ \\ 
	\hline 
	decay rate & $k^{\text{phys}} = 0.00066851 \, \left[1/s\right]$ &  $k^{\text{LB}} =  3.34255\times 10^{-7} \, \left[1/\delta_{t}\right]$\\ 
	\hline
	\end{tabular} 
    \label{tab:advectiondiffusionparameters}
\end{table}

\begin{table}[!ht]
    \centering
    \caption{\bf{Parameter values for the step profile advection test.}}

	\begin{tabular}{|c|c|c|c|c|}
	\hline 
	 & (a) & (b) & (c) & (d) \\ 
	\hline
	\hline
	$U^{\text{phys}}$ & $5000\,\left[\mu m/s\right]$ & $5000\,\left[\mu m/s\right]$  & $1000\,\left[\mu m/s\right]$  & $1000\,\left[\mu m/s\right]$  \\ 
	\hline 
	$D^{\text{phys}}$ & $122\,\left[\mu m^{2}/s\right]$ & $122\,\left[\mu m^{2}/s\right]$ & $122\,\left[\mu m^{2}/s\right]$ & $122\,\left[\mu m^{2}/s\right]$ \\ 
	\hline 
	$T^{\text{phys}}$ & $0.2\,\left[s\right]$ & $0.2\,\left[s\right]$ & $1.0\,\left[s\right]$ & $1.0\,\left[s\right]$ \\ 
	\hline 
	\hline
	$\delta_{x}$ & $5\,\left[\mu m\right]$ & $5\,\left[\mu m\right]$ & $5\,\left[\mu m\right]$ & $5\,\left[\mu m\right]$\\
	\hline
	$\delta_{t}$ & $0.0005\,\left[s\right]$ & $0.00005\,\left[s\right]$ & $0.0005\,\left[s\right]$ & $0.00005\,\left[s\right]$ \\
	\hline
	\hline
	$U^{\text{LB}}$ & $0.5 \, \left[\delta_{x}/\delta_{t}\right]$ & $0.05 \, \left[\delta_{x}/\delta_{t}\right]$ & $0.1 \, \left[\delta_{x}/\delta_{t}\right]$ & $0.01 \, \left[\delta_{x}/\delta_{t}\right]$ \\ 
	\hline 
	$D^{\text{LB}}$ & $0.00244\,\left[\delta_{x}^{2}/\delta_{t}\right]$ & $0.000244\,\left[\delta_{x}^{2}/\delta_{t}\right]$ & $0.00244\,\left[\delta_{x}^{2}/\delta_{t}\right]$ & $0.000244\,\left[\delta_{x}^{2}/\delta_{t}\right]$ \\ 
	\hline 
	\end{tabular} 
    \label{tab:shockparameters}
\end{table}

\end{document}